\documentclass[11pt]{article}
\pdfoutput=1
\usepackage{amsmath,epsf,amssymb,latexsym,array,color}
\usepackage{graphicx,wasysym}
\usepackage{cancel}

\usepackage{hyperref}
\def\be{\begin{equation}}
\def\ee{\end{equation}}
\def\ba{\begin{array}}
\def\ea{\end{array}}
\newcommand{\beq}{\begin{equation}}
\newcommand{\eeq}[1]{\label{#1}\end{equation}}
\newcommand{\bea}{\begin{eqnarray}}
\newcommand{\eea}[1]{\label{#1}\end{eqnarray}}

\begin{document}
%\title{Constraining Liberated Supergravity}
%\author{Hun Jang}
%\email{hun.jang@nyu.edu}
%\author{Massimo Porrati}
%\email{massimo.porrati@nyu.edu, hun.jang@nyu.edu}
%\affiliation{Center for Cosmology and Particle Physics\\ Department of Physics, New York University \\
%726 Broadway, New York NY 10003, USA}

%\maketitle

\begin{titlepage}

\hskip 1.5cm

\begin{center}
{\huge \bf{Inflation, Gravity Mediated Supersymmetry Breaking, and de Sitter Vacua in Supergravity with a K\"{a}hler-Invariant FI Term}}
\vskip 0.8cm  
{\bf \large Hun Jang and Massimo Porrati}  
\vskip 0.75cm

{\em Center for Cosmology and Particle Physics\\
	Department of Physics, New York University \\
	726 Broadway, New York, NY 10003, USA}
	
	\vspace{12pt}

\end{center}

\begin{abstract}
We use a new mechanism for generating a Fayet-Iliopoulos term in supergravity, which is not associated to an R symmetry, to 
construct a semi-realistic theory of slow-roll inflation for a theory with the same K\"ahler potential and superpotential as the KKLT
string background (without anti-D3 branes). In our model, supersymmetry must be broken at a high scale in a hidden sector 
to ensure that the cutoff of the effective field theory is above the Hubble scale of inflation. The
gravitino has a super-EeV mass and supersymmetry breaking is communicated to the observable sector through gravity mediation.
Some mass scales of the supersymmetry-breaking soft terms in the observable sector can be parametrically smaller than the SUSY 
breaking
 scale. If a string realization of the new FI term were found, our model could be the basis for a
 low energy effective supergravity description
 of realistic superstring models of inflation.

\end{abstract}

\vskip 1 cm
\vspace{24pt}
\end{titlepage}
\tableofcontents

\section{Introduction}
It is rather challenging to describe inflation, supersymmetry (SUSY) breaking, and de Sitter (dS) vacua in simple supergravity models and even more so in string theory. In string theory, the 
Kachru-Kallosh-Linde-Trivedi (KKLT) model \cite{KKLT} is a prototype
that can give de Sitter (dS) vacua, under certain assumptions about moduli stabilization. The effective field theory description of the
KKLT model is a supergravity with a no-scale K\"ahler potential for its volume modulus 
and with a superpotential that differs from its constant no-scale form because of two non-perturbative 
corrections\footnote{The corrections come from either Euclidean D3 branes in type IIB compactifications or from 
gaugino condensation due to D7 branes.}.  
The superpotential produces a supersymmetric Anti-de-Sitter (AdS) vacuum. In ref.~\cite{KKLT}, a mechanism was
 proposed for generating dS
 vacua through the addition of anti-D3 brane contributions to the superpotential, that uplifts the AdS vacuum to dS. While the 
additional correction by anti-D3 branes creates dS vacua, it also deforms the shape of the scalar potential creating a ``bump'' which 
gives rise to a moduli stabilization problem~\cite{KKLT}. As an attempt to improve on KKLT, Kachru, Kallosh, Linde, 
Maldacena, McAllister
 and Trivedi (KKLMMT) proposed a model that modifies KKLT by introducing a contribution arising from the anti-D3 tension in a highly warped compactifications~\cite{KKLMMT}.

Both models,  KKLT and the KKLMMT, contain anti-D3 branes, whose known effective field theory description uses nonlinear 
realizations of supersymmetry. The presence of  nonlinearly realized
 supersymmetry means that if supersymmetry is restored at energies below the string scale, $M_{string}$, then 
 the known description of 
 KKLT cannot accurately describe the whole energy range $E\lesssim M_{string}$\footnote{We assume $M_{string}< M_{pl}$, with 
 $M_{pl}$ the Planck scale.}. On the other hand, nothing in principle forbids the existence of {\em some} 
 effective field theory description even in that
 energy range, but such description must employ a linear realization of supersymmetry, which would necessarily employ only whole multiplets. 
A natural question to ask from an effective field theory point of view is whether such a description is possible. Said differently: does
a supergravity with the same K\"ahler potential and superpotential as KKLT exists, that breaks supersymmetry, gives rise to an 
inflationary potential and a dS post-inflationary vacuum, and is valid even at energy scales where supersymmetry is
restored?
We answer affirmatively to this question by adding to the KKLT effective theory a new Fayet-Iliopoulos (FI) term, in the form proposed 
by Antoniadis, Chatrabhuti, Isono and Knoops~\cite{acik}. 
We will show that this FI term also generates irrelevant operators that introduce a cutoff scale for the effective theory. We will
also show that differently from nonlinear realizations, this cutoff can be made larger than the supersymmetry breaking scale --and in
fact even larger than the Planck scale.     

Our construction begins with the observation that, in the absence of anti-D3 branes, the supergravity scalar potential of 
the KKLT model has a supersymmetric AdS vacuum and it becomes flat for large values of the volume modulus field. The flat direction could be used 
for constructing a viable model of inflation without eta problem, if the scalar potential minimum $V_0$ could be simply
 translated upward by a constant, 
$V_0\rightarrow V_0+ \mathrm{constant}$. This could happen if a constant positive FI term existed. This term was long thought to be 
forbidden in supergravity, since the only possible FI terms were thought to arise from gauging the R-symmetry~\cite{R-symm},
require an R-invariant superpotential~\cite{barb} and be subject to quantization conditions when the gauged R-symmetry is 
compact~\cite{Nathan}.
On the other hand, recently FI terms not associated with R-symmetry were proposed, starting 
with ref.~\cite{CFTV}.
We use here the K\"{a}hler-invariant FI term proposed in~\cite{acik} and we 
call it ``ACIK-FI'' to distinguish it from many other new FI terms
 suggested in the literature (for instance in~\cite{CFTV,oldACIK,AKK,Kuzenko}). 
To find an approximately flat potential for inflation and a dS post-inflationary vacuum, we add an ACKI-FI term  to
the  $\mathcal{N}=1$ supergravity describing the KKLT model without anti-D3 branes. 
 We must remark that a field-dependent generalization of the new K\"{a}hler-invariant FI term has 
 been introduced recently in ref.~\cite{ar}, which also studies the cosmological consequences of such a term.

In our model supersymmetry is spontaneously broken in a hidden sector at a very high but still sub-Planckian scale 
$M_{pl} \gg M_S \gg 10^{-15}M_{pl}$. 
We employ gravity mediation (see e.g. the review~\cite{GM}) to communicate the SUSY breaking to the observable sector, where supersymmetry breaking manifests itself through the existence of explicit soft 
SUSY breaking terms, characterized by an energy scale $M_{observable} \ll M_S$.
    The reason for a high $M_S$
  is that $M_S$ controls the magnitude of non-renormalizable fermionic terms that determine the cutoff of the effective theory. This is a
  feature that the ACIK-FI term shares with liberated supergravity (see e.g.~\cite{Liberated,jp1}).
   
 The purpose of our work is to find an effective field theory of inflation, de-Sitter moduli stabilization, and supersymmetry breaking as a
  cosmological application to the effective theory of KKLT of the ACIK-FI term proposed in~\cite{acik}. 
  The string theory origin of one particular type of the new FI 
 terms  has recently been investigated through a supersymmetric Born-Infeld action~\cite{CFT}, so it would be of clear 
 interest to study a possible string-theoretical origin of the ACIK-FI term. 

This paper is organized as follows. In Sec. 2 we show how to add an ACIK-FI term to the $\mathcal{N}=1$ supergravity 
effective theory of the KKLT model. Next we add matter, which we divide into a hidden sector and an observable sector. Supersymmetry 
is broken in the hidden sector and the SUSY breaking is communicated to the observable sector via gravity mediation.
  In Secs. 3 and 4 we probe the hidden-sector dynamics of our model. In Sec. 3, we construct a minimal supergravity model of 
 plateau-potential inflation --sometimes called in the literature ``Starobinsky'' or ``Higgs'' inflation-- with high scale SUSY breaking and dS vacua, using the results from Sec. 2. In Sec. 4, we explore the gravitino mass, which is very high, being
 well above the EeV-scale. We also study possible constraints on the ACIK-FI term by investigating the nonrenormalizable fermionic terms in the Lagrangian. These constraints can be satisfied if the gauge coupling for a certain U(1) necessarily present in our model
 is sufficiently small. They also lead to a hierarchy of energy scales. In Sec. 5 we 
  study the observable-sector dynamics of our model by computing its soft SUSY breaking terms. A few final observations are collected
  in Sec. 6.

\section{Adding a K\"{a}hler-invariant Fayet-Iliopoulous term to KKLT-type $\mathcal{N}=1$ supergravity}

In this section, we propose an $\mathcal{N}=1$ supergravity model that can describe the low energy effective field theory of inflation and moduli stabilization in KKLT-type backgrounds \cite{KKLT}. To do so, we first add an ACIK-FI term to an $\mathcal{N}=1$ supergravity that is compatible with the KKLT model.

In general, an ACIK-FI term can be introduced  into an $\mathcal{N}=1$ supergravity without requiring a gauged R-symmetry \cite{acik,ar}. In our proposal, we will introduce instead only 
an  ordinary U(1) symmetry (under which the superpotential is invariant) which will be gauged by a vector multiplet $V$. Inflation will come from the same potential as in the KKLT 
scenario. KKLT~\cite{KKLT} argues that in string theory some moduli can develop a non-perturbative superpotential of the form
\begin{eqnarray}
W = W_0 + A e^{-aT}, \label{kklt}
\end{eqnarray}
where $T$ is a ``volume'' modulus field, which is a chiral superfield, and $W_0,A$ are constants. 
For our construction it is sufficient to compute the component action of $\mathcal{N}=1$ supergravity characterized by the 
superpotential~\eqref{kklt} and by an ACIK-FI term. Notice that
 Antoniadis and Rondeau have recently studied  cosmological applications of generalized ACIK-FI terms by considering no-scale models
  with a constant superpotential $W=W_0$ \cite{ar}. Differently from that model, ours uses the KKLT-type superpotential~\eqref{kklt}. 

The key assumption that we will use is that both the volume modulus $T$ and the other matter fields that may exist in the 
superpotential are gauge-invariant under the U(1) that is used to introduce the ACIK-FI term. In this paper, we use superconformal 
tensor calculus \cite{cfgvnv} to calculate the action. 

The goal of this work is to find a modestly 
realistic minimal supergravity model of inflation with realistic moduli stabilization and supersymmetry
breaking pattern. The study of irrelevant operators generated by the ACIK-FI term will show that a low energy supersymmetry breaking
is incompatible with demanding that the cutoff for the effective field theory is higher than the Hubble constant during inflation. So, we 
take an alternative approach and break supersymmetry at a high scale in the hidden sector (as in {\it e.g.}~\cite{HighSUSYGUT} )
while keeping some of the scales of supersymmetry breaking interactions in the observable sector low~\cite{softSUSYbreaking}.

To do so, we first decompose matter into a hidden sector and an observable sector.  We will discuss them separately in Sections 3, 4
and 5. So we separate the field coordinates $y^A$ into 
\begin{eqnarray}
y^A \equiv (T,z^{\hat{I}}) \equiv (\{T,z^I\}^h,\{z^i\}^o),
\end{eqnarray}
where $\hat{I} \equiv (I,i)$ and $\{T,z^I\}^h$ are hidden-sector fields, while $\{z^i\}^o$ are the observable-sector ones. In addition to this, we write a generic superpotential $W$ as a sum of a hidden-sector term $W^h$ and observable-sector term $W^o$:
\begin{eqnarray}
W(y^A) \equiv W^h(T,z^I) + W^o(z^i).
\end{eqnarray}

We further assume that the hidden-sector superpotential carries a high energy scale compared to the observable-sector one. This 
implies that we decompose the F-term scalar potential into two different parts: a hidden sector F-term potential characterized by a
 high energy scale and observable-sector F-term potential containing only  low scale SUSY-breaking soft terms.

Next, to introduce an ACIK-FI term into our theory we suppose that the volume modulus multiplet $T$ and all observable-sector chiral matter multiplets $Z^i$ are neutral under an ordinary (non-R) 
U(1) gauge symmetry, while the hidden-sector chiral matter multiplets $z^I$ are charged, {\it i.e.} they transform as
\begin{eqnarray}
Z^i \rightarrow Z^i, \quad T \rightarrow T, \quad Z^I \rightarrow e^{-q_I\Omega}Z^I.
\end{eqnarray}
Here $q_I$ denote the U(1) gauge charges of the hidden-sector chiral multiplets $Z^I$ and $\Omega$ is the chiral multiplet containing 
in its lowest component the ordinary gauge parameter. We make these choices because we 
will introduce both a new FI term generated by a gauge vector multiplet and  a KKLT superpotential, which depends on  the volume modulus $T$ and must be gauge invariant under all gauge symmetries.

The superconformal action of the ACIK-FI term \cite{acik,ar} is defined by
\begin{eqnarray}
\mathcal{L}_{\textrm{NEW FI}} \equiv - \xi \left[(S_0\bar{S}_0e^{-K(Ze^{qV},\bar{Z})})^{-3}\frac{(\mathcal{W}_{\alpha}(V)\mathcal{W}^{\alpha}(V))(\bar{\mathcal{W}}_{\dot{\alpha}}(V)\bar{\mathcal{W}}^{\dot{\alpha}}(V))}{T(\bar{w}^2)\bar{T}(w^2)}(V)_D\right]_D,\label{newFIterm}
\end{eqnarray}
and the corresponding superconformal action of $\mathcal{N}=1$ supergravity with superpotential~\eqref{kklt}  and the 
new the FI term is 
\begin{eqnarray}
\mathcal{L} &=& -3[S_0\Bar{S}_0e^{-K(Ze^{qV},\bar{Z})/3}]_D + [S_0^3W(Z,Z')]_F + \frac{1}{2g^2}[
\mathcal{W}_{\alpha}(V)\mathcal{W}^{\alpha}(V)]_F + c.c. \nonumber\\
&&- \xi \left[(S_0\bar{S}_0e^{-K(Ze^{qV},\bar{Z})})^{-3}\frac{(\mathcal{W}_{\alpha}(V)\mathcal{W}^{\alpha}(V))(\bar{\mathcal{W}}_{\dot{\alpha}}(V)\bar{\mathcal{W}}^{\dot{\alpha}}(V))}{T(\bar{w}^2)\bar{T}(w^2)}(V)_D\right]_D. \label{newFI2}
\end{eqnarray}
In Eqs.~(\ref{newFIterm},\ref{newFI2}) $S_0$ is the conformal compensator with Weyl/chiral weights (1,1); $Z^A = (T,Z^I;Z^i)$ and $V$ are chiral matter and vector multiplets with weights $(0,0)$; $K(Ze^{qV},\bar{Z})$ is a K\"{a}hler potential gauged by a vector multiplet $V$; $W(Z,Z')$ is a 
superpotential;
$\mathcal{W}_{\alpha}(V)$ is the field strength of the vector multiplet $V$; $\xi$ is the constant coefficient of ACIK-FI term; 
$w^2 \equiv \frac{\mathcal{W}_{\alpha}(V)\mathcal{W}^{\alpha}(V)}{(S_0\bar{S}_0e^{-K(Z,\bar{Z})})^{2}}$ and $\bar{w}^2 \equiv \frac{\bar{\mathcal{W}}_{\dot{\alpha}}(V)\bar{\mathcal{W}}^{\dot{\alpha}}(V)}{(S_0\bar{S}_0e^{-K(Z,\bar{Z})})^{2}}$ are composite 
multiplets, $T(X), \bar{T}(X)$ are chiral projectors, and $(V)_D$ is a real multiplet, whose lowest component is the auxiliary 
field $D$ of the vector multiplet $V$.  

Next, we write the following K\"ahler potential, invariant under the same U(1) that generates the ACIK-FI term
\begin{eqnarray}
K(Z^Ae^{qV},\bar{Z}^{\bar{A}})\equiv -3\ln[T+\Bar{T} - \Phi(Z^Ie^{qV},\bar{Z}^{\bar{I}}; Z^i,\bar{Z}^{\bar{i}})/3] ,
\end{eqnarray}
where $\Phi$ is a real function of the matter multiplets $Z^i,Z^I$ and the two terms in the superpotential 
$W \equiv W^h + W^o$ are the hidden-sector term
\begin{eqnarray}
W^h(T) \equiv W_0 + Ae^{-aT}
\end{eqnarray}
and the observable-sector superpotential 
\begin{eqnarray}
W^o(Z^i) \equiv B_0 + S_iZ^i + M_{ij}Z^iZ^j + Y_{ijk}Z^iZ^jZ^k + \cdots,
\end{eqnarray}
where $B_0,S_i,M_{ij},Y_{ijk}$ are constant coefficients.
We will choose $\Phi$ to be sum of a term containing only hidden-sector fields and one containing only those of the
observable sector 
\begin{equation}
\Phi=\Phi^h(Z^Ie^{qV},\bar{Z}^{\bar{I}})+ \Phi^o(Z^i,\bar{Z}^{\bar{i}}) . \label{factor}
\end{equation}

The supergravity G-function corresponding to our model is then 
\begin{eqnarray}
G(y^A,\bar{y}^{\bar{A}}) \equiv K(y^A,\bar{y}^{\bar{A}}) + \ln|W(y^A)|^2 = -3\ln[T+\bar{T}-\frac{\Phi(z^{\hat{I}},\bar{z}^{\bar{\hat{I}}})}{3}] + \ln|W^h(T,z^I)+W^o(z^i)|^2.
\end{eqnarray}
The F-term supergravity scalar potential is given by the formula $V_F \equiv e^G(G_AG^{A\bar{B}}G_{\bar{B}}-3)$, which in our case 
reads
\begin{eqnarray}
V_F &=& 
-\frac{1}{X^{2}}[(W^h+W^o)\bar{W}^h_{\bar{T}}+(\bar{W}^h+\bar{W}^o)W^h_{T}]
\nonumber\\&&
+\frac{1}{3}\frac{|W^h_T|^2}{X^{2}}+\frac{1}{9}\frac{|W^h_T|^2}{X^{2}}[\Phi_{I}\Phi^{I\bar{J}}\Phi_{\bar{J}}+\Phi_{i}\Phi^{i\bar{j}}\Phi_{\bar{j}}] \nonumber\\&&
+ \frac{1}{3}\frac{1}{X^{2}}
[W^h_T(\Phi_{I}\Phi^{I\bar{J}}\bar{W}^h_{\bar{J}}+\Phi_{i}\Phi^{i\bar{j}}\bar{W}^o_{\bar{j}})
+\bar{W}^h_{\bar{T}}(W^h_{I}\Phi^{I\bar{J}}\Phi_{\bar{J}}+W^o_{i}\Phi^{i\bar{j}}\Phi_{\bar{j}})] \nonumber\\&&
+\frac{1}{X^{2}}[W^h_I\Phi^{I\bar{J}}\bar{W}^h_{\bar{J}}+W^o_i\Phi^{i\bar{j}}\bar{W}^o_{\bar{j}}]. \label{scal-pot}
\end{eqnarray}

When matter scalars are charged under a gauge group there exists also a D-term contribution to the scalar potential, $V_D$.
 In our model, we find it to be
\beq
V_D = \frac{1}{2}g^2  \Big(\xi+\sum_I(q_Iz^IG_I+q_I\bar{z}^{\bar{I}}G_{\bar{I}})\Big)^2 = 
\frac{1}{2}g^2   \Big(\xi+\frac{q_Iz^I\Phi_I+q_I\bar{z}^{\bar{I}}\Phi_{\bar{I}}}{X}\Big)^2,
\eeq{D-term}
where $X \equiv T+\bar{T}-\Phi/3$, $g$ is the gauge coupling constant and $\xi$ is the ACIK-FI constant. Remember that only hidden-sector chiral matter multiplets are charged under the U(1). 
The scalar potential is the sum of two terms. One, $V_h$  contains the D-term contribution and the F-term potential of the hidden 
sector, depends on the high mass scale $M_S$ and is $O(H^2M_{pl}^2)$ during inflation; the other, $V_{soft}$ contains the observable sector scalars and depends only on low mass scales: 
\begin{eqnarray}
V = V_h + V_{soft},
\end{eqnarray}
where
\begin{eqnarray}
V_h &\equiv& V_D - \frac{W^h_T\bar{W}^h+\bar{W}^h_{\bar{T}}W^h}{X^{2}} + \frac{|W^h_T|^2}{3X^{2}}\Big(X+\frac{1}{3}\Phi_I\Phi^{I\bar{J}}\Phi_{\bar{J}}\Big) \nonumber\\
&& + \frac{1}{3}\frac{1}{X^{2}}
[W^h_T\Phi_{I}\Phi^{I\bar{J}}\bar{W}^h_{\bar{J}}
+\bar{W}^h_{\bar{T}}W^h_{I}\Phi^{I\bar{J}}\Phi_{\bar{J}}]+\frac{1}{X^{2}}W^h_I\Phi^{I\bar{J}}\bar{W}^h_{\bar{J}} ,\\
V_{soft} &\equiv&  
-\frac{1}{X^{2}}[W^o\bar{W}^h_{\bar{T}}+\bar{W}^oW^h_{T}]
+\frac{1}{9}\frac{|W^h_T|^2}{X^{2}}\Phi_{i}\Phi^{i\bar{j}}\Phi_{\bar{j}} \nonumber\\&&
+ \frac{1}{3}\frac{1}{X^{2}}
[W^h_T\Phi_{i}\Phi^{i\bar{j}}\bar{W}^o_{\bar{j}}
+\bar{W}^h_{\bar{T}}W^o_{i}\Phi^{i\bar{j}}\Phi_{\bar{j}}] 
+\frac{1}{X^{2}}W^o_i\Phi^{i\bar{j}}\bar{W}^o_{\bar{j}}.
\end{eqnarray}

\section{Hidden sector dynamics 1: a minimal supergravity model of inflation, high-scale supersymmetry breaking, and de Sitter vacua}

In this section, we explore a minimal supergravity model of high-scale supersymmetry breaking and plateau-potential inflation through gravity mediation and no-scale K\"ahler potential. We investigate first the hidden sector dynamics. We have assumed that the hidden-sector potential depends on a high energy scale and dominates over the observable-sector one. Hence, it is reasonable to minimize the hidden-sector potential first. 
Let us compute now the F-term potential in the hidden sector. Recalling that the KKLT superpotential is
\begin{eqnarray}
W^h(T) \equiv W_0 + Ae^{-aT},
\end{eqnarray}
and redefining $W_0 \equiv -cA$, we rewrite it as
\begin{eqnarray}
W^h(T) = A(e^{-aT}-c),
\end{eqnarray}
where $a,c,A$ are positive constants. Note that $W^h_I = \partial W^h/\partial z^I = 0$.

Since we defined $X \equiv T+\bar{T}-\Phi/3$, the KKLT superpotential gives 
\begin{eqnarray}
&& W^h_T = -aAe^{-aT},\quad |W^h_T|^2 = a^2A^2 e^{-a(T+\bar{T})}= a^2A^2e^{-a(X+\Phi/3)}, 
\end{eqnarray}
\begin{eqnarray}
W^h_T\bar{W}^h+\bar{W}^h_{\bar{T}}W^h &=& -aA^2e^{-aT}(e^{-a\bar{T}}-c) -aA^2e^{-a\bar{T}}(e^{-aT}-c) \nonumber\\
&=& -2aA^2e^{-a(T+\bar{T})}+aA^2c(e^{-aT}+e^{-a\bar{T}}) \nonumber\\
&=& -2aA^2e^{-a(X+\Phi/3)}+aA^2c(e^{-a(\textrm{Re}T+i\textrm{Im}T)}+e^{-a(\textrm{Re}T-i\textrm{Im}T)}) \nonumber\\
&=& -2aA^2e^{-a(X+\Phi/3)}+2aA^2c e^{-a\textrm{Re}T} \cos(a\textrm{Im}T)
\nonumber\\
&=& -2aA^2e^{-a(X+\Phi/3)}+2acA^2 e^{-a(X+\Phi/3)/2} \cos(a\textrm{Im}T)
\end{eqnarray}
\begin{eqnarray}
|W^h|^2 &=& A^2|e^{-aT}-c|^2 = A^2 (e^{-aT}-c)(e^{-a\bar{T}}-c)= A^2(e^{-a(T+\bar{T})}-c(e^{-aT}+e^{-a\bar{T}})+c^2) \nonumber\\
&=& A^2 (e^{-a(X+\Phi/3)}-2ce^{-a(X+\Phi/3)/2} \cos(a\textrm{Im}T)+c^2).
\end{eqnarray}
Here we have used the following transformation from the complex coordinate $T$ to two real coordinates $X,\textrm{Im}T$:
\begin{eqnarray}
T = \textrm{Re}T+i\textrm{Im}T = \frac{1}{2}\Big(X+\frac{\Phi}{3}\Big) + i\textrm{Im}T, 
\end{eqnarray}
which gives $e^{-aT} = e^{-\frac{a}{2}(X+\frac{\Phi}{3})}e^{-a i\textrm{Im}T}$. Remember that $X \equiv T+\bar{T}-\Phi/3$.

Then, since $W^h_I=0$, the corresponding hidden-sector F-term scalar potential is given by
\begin{eqnarray}
V^h_F &=& - \frac{W^h_T\bar{W}^h+\bar{W}^h_{\bar{T}}W^h}{X^{2}} + \frac{|W^h_T|^2}{3X^{2}}\Big(X+\frac{1}{3}\Phi_I\Phi^{I\bar{J}}\Phi_{\bar{J}}\Big) \nonumber\\
&=& - \frac{1}{X^2}\bigg(-2aA^2e^{-a(X+\Phi/3)}+2acA^2 e^{-a(X+\Phi/3)/2} \cos(a\textrm{Im}T)\bigg)
\nonumber\\&&+ \frac{1}{3X^{2}}\Big(X+\frac{1}{3}\Phi_I\Phi^{I\bar{J}}\Phi_{\bar{J}}\Big) a^2A^2 e^{-a(X+\Phi/3)}.
\end{eqnarray}

Since we assume that the D-term potential belongs to the hidden sector, the hidden-sector total scalar potential can be written as
\begin{eqnarray}
V_h &=& V_D+V_F^h  \nonumber\\
&=& \frac{1}{2}g^2   \Big(\xi+\frac{q_Iz^I\Phi_I+q_I\bar{z}^{\bar{I}}\Phi_{\bar{I}}}{X}\Big)^2 - \frac{1}{X^2}\bigg(-2aA^2e^{-a(X+\Phi/3)}+2acA^2 e^{-a(X+\Phi/3)/2} \cos(a\textrm{Im}T)\bigg)
\nonumber\\&&+ \frac{1}{3X^{2}}\Big(X+\frac{1}{3}\Phi_I\Phi^{I\bar{J}}\Phi_{\bar{J}}\Big) a^2A^2 e^{-a(X+\Phi/3)}.
\end{eqnarray}
We define the SUSY breaking scale $M_S$ in terms of the scalar potential $V_h$ and the gravitino mass $m_{3/2}$ as
\begin{eqnarray}
V_+ \equiv M_S^4 = V_h+3m_{3/2}^2 = V_h + \frac{3}{X^3}A^2 (e^{-a(X+\Phi/3)}-2ce^{-a(X+\Phi/3)/2} \cos(a\textrm{Im}T)+c^2).
\end{eqnarray}

To investigate the moduli stabilization, we identify the canonically normalized fields by inspection of the kinetic terms, which are given by
\begin{eqnarray}
\mathcal{L}_K &=& \frac{\Phi_{\hat{I}\bar{\hat{J}}}}{X}g_{\mu\nu} D_{\mu}z^{\hat{I}}D^{\nu}\bar{z}^{\bar{\hat{J}}} + \frac{3}{4X^2}g_{\mu\nu}\partial_{\mu}X\partial_{\nu}X
\nonumber\\
&&+\frac{3}{X^2}g_{\mu\nu}[\partial_{\mu}\textrm{Im}T-(\textrm{Im}D_{\mu}z^{\hat{I}}\Phi_{\hat{I}}/3)][\partial_{\nu}\textrm{Im}T-(\textrm{Im}D_{\nu}z^{\hat{I}}\Phi_{\hat{I}}/3)] ,
\end{eqnarray}
where $D_{\mu} \equiv \partial_{\mu} - iq_{\hat{I}}A_{\mu}$ is the U(1) gauge covariant derivative for the matter multiplets $z^{\hat{I}} = (z^I,z^i)$ with gauge charge $q_{\hat{I}}=(q_I\neq0,q_i=0)$, and $A_{\mu}$ is the corresponding gauge field.

After performing another field redefinition $X \equiv e^{\sqrt{2/3}\phi}$, we find 
\begin{eqnarray}
\mathcal{L}_K &=& \Phi_{I\bar{J}}e^{-\sqrt{2/3}\phi}g_{\mu\nu} D_{\mu}z^ID^{\nu}\bar{z}^{\bar{J}} + \frac{1}{2}g_{\mu\nu}\partial_{\mu}\phi\partial_{\nu}\phi
\nonumber\\
&&+3e^{-2\sqrt{2/3}\phi}g_{\mu\nu}[\partial_{\mu}\textrm{Im}T-(\textrm{Im}D_{\mu}z^I\Phi_I/3)][\partial_{\nu}\textrm{Im}T-(\textrm{Im}D_{\nu}z^I\Phi_I/3)].\label{kinetic_term}
\end{eqnarray}
Notice that $\phi$ is canonically normalized, while the other fields $z^I, \textrm{Im}T$ are so only when $\phi$ is small.

Now, let us investigate the scalar potential vacuum. First of all, we find the minimum with respect to the matter scalars $z^{\hat{I}}$, 
\begin{eqnarray}
\frac{\partial V}{\partial z^{\hat{I}}} = 0 \implies \Phi_{{\hat{I}}} = 0. 
\end{eqnarray}
If we choose a real function such that $\Phi_{\hat{I}}=0$ implies $\Phi =0$ together with $z^{\hat{I}}=0$ then at this vacuum the scalar potential becomes~\footnote{The observable-sector superpotential $W^o$ can shift the VEVs
of the scalars in the observable sector $z^i$, but since those VEVs must be in any case small compared to $H$ and $M_{pl}$ we can
approximately set $z^i=0$. Moreover, in our toy example in Section 5 we will choose a superpotential that indeedgives a 
minimum at $z^i=0$.}
\begin{eqnarray}
V_h &=& \frac{1}{2}g^2\xi^2 - \frac{1}{X^2}\bigg(-2aA^2e^{-aX}+2acA^2 e^{-aX/2} \cos(a\textrm{Im}T)\bigg)+ \frac{1}{3X} a^2A^2 e^{-aX}.
\end{eqnarray}
Next, we consider the vacuum with respect to the $\textrm{Im}T$ field. We find the vacuum at $a\textrm{Im}T=n\pi$, where $n$ is an 
even integer, leading to $\cos(a\textrm{Im}T)=1$\footnote{When $\cos(a\textrm{Im}T)=1$, the second derivative of the potential can be positive, which means that the stationary point is a minimum.} and
\begin{eqnarray}
V_h &=& \frac{1}{2}g^2\xi^2 +\frac{2aA^2}{X^2}e^{-aX}- \frac{2acA^2}{X^2} e^{-aX/2}+ \frac{a^2A^2}{3X}  e^{-aX}.
\end{eqnarray}

Next, let us find the vacuum with respect to the $\phi$ field. Recalling that  $X=e^{\sqrt{2/3}\phi}$, calling 
$\langle \phi \rangle$ the vacuum expectation value of $\phi$ and setting
$\phi=\left<\phi\right> = \sqrt{\frac{3}{2}}\ln \left<X\right> = \sqrt{\frac{3}{2}}\ln x$, where  $X = \left<X\right> \equiv x$, we have
\begin{eqnarray}
\frac{\partial V_h}{\partial \phi}\bigg|_{\phi=\left<\phi\right>} = \frac{\partial V_h}{\partial X}\bigg|_{X=x}\frac{\partial X}{\partial \phi} \bigg|_{\phi=\left<\phi\right>}= 0 \implies \frac{\partial V_h}{\partial X}\bigg|_{X=x} = 0,
\end{eqnarray}
which gives 
\begin{eqnarray}
\frac{\partial V_h}{\partial X}\bigg|_{X=x}=-\frac{4aA^2}{x^3} e^{-ax} - \frac{2a^2A^2}{x^2} e^{-ax} + \frac{4acA^2}{x^3}e^{-ax/2} + \frac{a^2cA^2}{x^2}e^{-ax/2} - \frac{a^2A^2}{3x^2}e^{-ax} - \frac{a^3A^2}{3x}e^{-ax} = 0. \nonumber\\
{}
\end{eqnarray}
At first glance, this equation seems a little complicated, but after a short calculation, we can obtain the following simple relation
\begin{eqnarray}
\frac{\partial V_h}{\partial X}\bigg|_{X=x} = 0 \implies c = \Big(1+\frac{ax}{3}\Big)e^{-ax/2}.
\end{eqnarray}

Inserting the value of $c$ into $V_h$, we obtain the following equation
\begin{eqnarray}
V_h &=& \frac{1}{2}g^2\xi^2 +\frac{2aA^2}{X^2}e^{-aX}- \frac{2aA^2}{X^2}\Big(1+\frac{ax}{3}\Big)e^{-ax/2} e^{-aX/2}+ \frac{a^2A^2}{3X}  e^{-aX},
\end{eqnarray}
where $X = e^{\sqrt{2/3}\phi}$.

Then, the vacuum energy at $X=x$ is given by
\begin{eqnarray}
V_h|_{X=x} = \frac{1}{2}g^2\xi^2 - \frac{a^2A^2e^{-ax}}{3x} \equiv \Lambda,
\end{eqnarray}
where $\Lambda$ is defined to be the post-inflationary cosmological constant, and the SUSY breaking scale is given by
\begin{eqnarray}
V_+|_{X=x} &=& V_h|_{X=x} + \frac{3}{X^3}A^2 (e^{-aX}-2ce^{-aX/2}+c^2)\bigg|_{X=x} \nonumber\\
&=& \Lambda + \frac{3}{x^3}A^2 (e^{-ax/2}-c)^2 = \Lambda + \frac{3A^2}{x^3} \frac{a^2x^2e^{-ax}}{9}\nonumber\\
&=& \Lambda + \frac{a^2A^2e^{-ax}}{3x} = \frac{1}{2}g^2\xi^2 \equiv M_S^4 ,
\end{eqnarray}
where $M_S$ is by definition the SUSY breaking scale.  
We can set $\Lambda$ to any value we wish, in particular we can choose it to be 
$\Lambda\sim 10^{-120}$. 

Here, we point out that the term $ \frac{1}{2}g^2\xi^2$ governs the magnitude of the total scalar potential, and simultaneously controls the scale of spontaneously supersymmetry breaking. Hence, if we want that the scalar potential describes inflation, we need to impose 
\begin{eqnarray}
M_S^4 =  \frac{1}{2}g^2\xi^2 \overset{!}{=} H^2M_{pl}^2 \equiv M_I^4,
\end{eqnarray}
where $H$ is the Hubble parameter, and $M_I$ is defined to be the mass scale of inflation. 

We then identify 
\begin{eqnarray}
A = \sqrt{\frac{3x(M_I^4-\Lambda)e^{ax}}{a^2}}, \qquad W_0 = -cA = -\Big(1+\frac{ax}{3}\Big)\sqrt{\frac{3x(M_I^4-\Lambda)}{a^2}}.
\end{eqnarray}

Substituting the above parameters $A,W_0,M_I$ into the hidden-sector potential, we can obtain a plateau inflation potential 
\begin{eqnarray}
V_h = M_I^4 - (M_I^4-\Lambda)x \bigg[ \frac{6e^{-a(X-x)/2}}{aX^2}\left(1-e^{-a(X-x)/2}+\frac{ax}{3}\right)-\frac{e^{-a(X-x)}}{X}  \bigg],
\end{eqnarray}
where $X = e^{\sqrt{2/3}\phi}$ and $\phi$ is defined to be the inflaton. Notice that the inflaton mass after inflation is of order of the 
Hubble scale, {\it i.e.} $m_{\phi}^2 \sim H^2 = 10^{-10}M_{pl}^2$.

We note that the hidden-sector scalar potential $V_h$ has a plateau, so it is of HI type (in the notations of ref.~\cite{LightScalar}).
Furthermore, it depends only on four parameters, which are: the vacuum expectation value of $X$ ({\it i.e.} $x \equiv \left<X\right>$); the KKLT parameter $a$ in the superpotential, which will be determined according to the type of the nonperturbative correction we choose\footnote{ For example, if we consider a nonperturbative correction due to gaugino condensation, then we find $a = \frac{2\pi}{N_c}$ for a non-abelian gauge group $SU(N_c)$ where $N_c$ is interpreted as the number of coincident D7 branes being stacked \cite{KKLT}.}; the inflation scale $M_I$, and the post-inflationary cosmological constant $\Lambda$. 
At $X=x$, the potential indeed reduces to the post-inflationary cosmological constant. As an additional remark, we observe that for fixed $x,M_I,\Lambda$ inflation ends earlier when $a$ is smaller. When the  
nonperturbative corrections to the KKLT superportential come from gaugino condensation~\cite{KKLT}, a smaller parameter $a$
corresponds to more D7 branes being stacked.

\section{Hidden sector dynamics 2: super-EeV gravitino mass, weak gauge coupling, and a hierarchy of energy scales}

In this section, we investigate some physical implications that can be obtained from our model. 
First of all, let us find the gravitino mass after inflation, which is
 generated by the high-scale SUSY breaking in the hidden sector. It is given by
\begin{eqnarray}
m_{3/2}^2 &=& e^G = \frac{|W^h|^2}{X^3} = \frac{A^2}{X^3}(e^{-a(X+\Phi/3)}-2ce^{-a(X+\Phi/3)/2} \cos(a\textrm{Im}T)+c^2)\bigg|_{z^I=0,a\textrm{Im}T=0,X=x} \nonumber\\
&=& \frac{3x(M_I^4-\Lambda)e^{ax}}{a^2x^3}(e^{-ax}-2ce^{-ax/2}+c^2)
=\frac{3x(M_I^4-\Lambda)e^{ax}}{a^2x^3}(c-e^{-ax/2})^2 \nonumber\\
&=& \frac{(M_I^4-\Lambda)}{3} \implies m_{3/2} \approx \frac{H}{\sqrt{3}}  = 10^{-6}M_{pl} \sim 10^{12}~\textrm{GeV} = 10^3 ~\textrm{EeV},
\end{eqnarray}
which is compatible with the case of EeV-scale gravitino cold dark matter candidates. This is not surprising because we are considering
 the same high-scale supersymmetry breaking scale as in \cite{EeVGravitino,Inf_HighSUSY_EeVGrav,HeavyGravitino,GDM}, where
 that scenario was proposed. The possibility of direct detection for such heavy dark matter candidates has recently been studied in ref. \cite{HDM}. Notice that in our model the gravitino mass is always $\mathcal{O}(H)$, irrespective of the ultraviolet cutoff.

Next, we explore possible constraints on the FI term by analyzing the fermionic nonrenormalizable interaction terms that are 
induced by such term~\eqref{newFIterm}. Schematically the general fermionic terms have the form
\begin{eqnarray}
 \mathcal{L}_F \supset \xi M_{pl}^{4m+2}D^{-2m-4+p} \mathcal{O}_F^{(10-2p)},\label{FermionicTerms}
\end{eqnarray}
where $\xi$ is the dimensionless ACIK-FI constant; $\mathcal{O}^{(\delta)}_F$ is an effective field operator of dimension $\delta$,
which does not contain any power of $D$;  $m$ is the total order of derivatives with respect to the 
the composite chiral fields  $T(\bar{w}^2)$ and $\bar{T}(w^2)$ defined after Eq.~\eqref{newFI2},
 and $p=0,1$. Detailed calculations will be given in \cite{jp20}, here we will only briefly summarize the main points.

To evaluate the fermionic terms we need to solve for the auxiliary field $D$. Equation~\eqref{newFIterm} gives the following 
Lagrangian for them
\begin{eqnarray}
\mathcal{L}_{\textrm{aux D}}= \frac{1}{2}D^2 -i(G_ik^i-G_{\bar{i}}k^{\bar{i}})D - \xi D \equiv \frac{1}{2}D^2-(\xi'+\xi)D,
\label{d-term}
\end{eqnarray}
where $\xi$ is the new FI constant while  $\xi' \equiv i(G_iK^i-G_{\bar{i}}K^{\bar{i}})$ is the standard field-dependent linear
term in $D$. It is written in terms of the Killing vector $K=K^i\partial_i$ giving the action of our U(1) gauge symmetry on the scalar fields.
$G$ is the standard supergravity G-function~\cite{cfgvnv}. 
Restoring the dependence on the gauge coupling constant $g$ we have
\begin{eqnarray}
\mathcal{L}_{\textrm{aux D}}= \frac{1}{2g^2 }D^2 -\xi'D - \xi  D = \frac{1}{2g^2 }D^2-(\xi'+\xi)D.
\end{eqnarray}
After solving the equation of motion for $D$, we find the solution 
\begin{eqnarray}
 D = g^2 M_{pl}^2(\xi+\xi') ~(= g^2(\xi+\xi')~\textrm{when}~M_{pl}=1).
\end{eqnarray}
Plugging this solution into the fermionic terms in Eq.~\eqref{FermionicTerms}  we obtain
\begin{eqnarray}
  \mathcal{L}_F &\supset& \xi M_{pl}^{4m+2}(g^2M_{pl}^2(\xi+\xi'))^{-2m-4+p} \mathcal{O}_F^{(10-2p)} \nonumber\\
  &=& \xi(g^2(\xi+\xi'))^{-2m-4+p} M_{pl}^{-6+2p} \mathcal{O}_F^{(10-2p)}.
\end{eqnarray}
The fermionic nonrenormalizable interactions generated by the ACIK-FI term introduce a strong coupling scale that sets the limit of validity of the effective field theory description. If we demand that the theory is
valid up to some cutoff scale $\Lambda_{cut}$, we find the folloiwing constraint on the ACIK-FI term:
\begin{eqnarray}
\xi(g^2(\xi+\xi'))^{-2m-4+p} \lesssim \left(\frac{M_{pl}}{\Lambda_{cut}}\right)^{6-2p}.
\end{eqnarray}

We must also examine the constraints on the post-inflation vacuum, that is the true vacuum, in which $\xi'=0$. 
In this case, we obtain 
\begin{eqnarray}
(g^{-1})^2 \left(\frac{\Lambda_{cut}}{M_{pl}}\right)^{6-2p} < (g^{-1})^{\dfrac{2(2m+4-p)}{(2m+3-p)}} \left(\frac{\Lambda_{cut}}{M_{pl}}\right)^{6-2p} \lesssim \xi .
\end{eqnarray}
This inequality reduces to the following: for all  $\Lambda_{cut} \leq M_{pl}$, we obtain 
\begin{eqnarray}
g^{-2} \left(\frac{\Lambda_{cut}}{M_{pl}}\right)^{4} \leq \xi .
\end{eqnarray}
Now we are ready to ask how does this constraint affect our supergravity model of inflation. To answer this, let us get back to the definition of the inflation scale (restoring the mass dimension)
\begin{eqnarray}
M_I^4 = \frac{1}{2}g^2\xi^2M_{pl}^4 \implies \xi = \sqrt{2} \frac{M_I^2}{M_{pl}^2}g^{-1} \sim 10^{-5}g^{-1}.
\end{eqnarray}
Inserting this equation into the constraints, we find that for $\Lambda_{cut} \leq M_{pl}$:
\begin{eqnarray}
 10^5\left(\frac{\Lambda_{cut}}{M_{pl}}\right)^{4} \leq g \label{Cutoff_GaugeCoupling_Constraint}
\end{eqnarray}
We see from this equation that it is easy to obtain $g \lesssim 1$. Let us define $\Lambda_{cut} \equiv 10^{k}M_{pl}$ where $k \in \mathbb{R}$ and plug this into the constraint in Eq.~\eqref{Cutoff_GaugeCoupling_Constraint}. Then, we have
\begin{eqnarray}
    10^{5+4k} \leq g.
\end{eqnarray}
If we demand a small gauge coupling such that $g\lesssim 1$, then the constraint reduces to
\begin{eqnarray}
      10^{5+4k} \leq g \lesssim 1 \implies k \lesssim -1.25 \implies \Lambda_{cut} \lesssim 10^{-1.25} M_{pl} < M_{pl}. \label{Cutoff_constraint}
\end{eqnarray}
Therefore, we note that requiring a small gauge coupling such that $g \lesssim 1$ guarantees that the cutoff scale $\Lambda_{cut}$ is lower than the Planck scale $M_{pl}$ and enables us to choose any sub-Planckian cutoff scale up to the upper bound in Eq.~\eqref{Cutoff_constraint}.  

As an example, if we assume that the cutoff of our theory is given
 by a Grand Unified (GUT) scale ({\it i.e.} $\Lambda_{cut} \sim 10^{-2}M_{pl} \gg H \approx 10^{-5}M_{pl}$), then we find that the gauge coupling must  only obey $10^{-3} \leq  g$, so it can easily obey $g  \lesssim 1$. Remember that our SUSY breaking scale was given by $M_S = M_I = \sqrt{HM_{pl}} \sim 10^{-2.5}M_{pl}$, which is slightly below the GUT-scale cutoff, {\it i.e.} $M_S < \Lambda_{cut}=\Lambda_{GUT}=10^{-2}M_{pl}$. Consequently, we have to consider an effective theory with the following hierarchy of scales: $H \ll M_S \lesssim \Lambda_{cut} = \Lambda_{GUT} < M_{pl}$, to ensure that the  gauge coupling constant obeys $ O(10^{-3}) \lesssim g \lesssim 1$. We may also set the cutoff at the string scale $M_{\textrm{string}} \sim 10^{-3} M_{pl}$. In this case, the hierarchy of mass scales is given by $H \ll \Lambda_{cut}=M_{\textrm{string}} < M_S < M_{pl}$.

\section{Observable sector dynamics: low scale soft supersymmetry breaking interactions}

In this section we investigate the mass scales of the soft supersymmetry-breaking interactions in the observable sector. We
need to find under which conditions our model could be phenomenologically realistic. A full investigation of the detailed structure of the 
soft interactions in the observable sector requires a study that goes beyond the scope of this work, so here we will limit ourselves to
general remarks and a coarse-grained analysis of necessary conditions for the viability of our model. We focus our
analysis on the soft masses. 

Restoring the mass dimension (so that the $T,z^i$ have  canonical mass dimension 1), the soft-term potential becomes
\begin{eqnarray}
V_{soft} &\equiv&  
-\frac{1}{M_{pl}X^{2}}[W^o\bar{W}^h_{\bar{T}}+\bar{W}^oW^h_{T}]
+\frac{1}{9}\frac{|W^h_T|^2}{M_{pl}^2X^{2}}\Phi_{i}\Phi^{i\bar{j}}\Phi_{\bar{j}} \nonumber\\&&
+ \frac{1}{3}\frac{1}{M_{pl}X^{2}}
[W^h_T\Phi_{i}\Phi^{i\bar{j}}\bar{W}^o_{\bar{j}}
+\bar{W}^h_{\bar{T}}W^o_{i}\Phi^{i\bar{j}}\Phi_{\bar{j}}] 
+\frac{1}{X^{2}}W^o_i\Phi^{i\bar{j}}\bar{W}^o_{\bar{j}}.
\end{eqnarray}
This formula is obtained by taking the following low-energy limit:  $F^h, M_{pl} \rightarrow \infty$ (where $F^h$ are the hidden-sector auxiliary F-term fields) while $m_{3/2}$=constant~\cite{GM}. Elegant examples of gravity mediation and soft SUSY breaking are simply explained in {\it e.g.}~\cite{fvp}. 

In addition, the hidden-sector superpotential can be written as 
\begin{eqnarray}
W^h &=& A(e^{-aT/M_{pl}}-c)=M_{pl}\sqrt{\frac{3x(M_I^4-\Lambda)e^{ax}}{a^2}}(e^{-aT/M_{pl}}-(1+ax/3)e^{-a/2}) \nonumber\\
&=& M_{pl}\sqrt{\frac{3x(M_I^4-\Lambda)e^{ax}}{a^2}}(e^{-a(X+\Phi/3M_{pl}^2)/2}e^{-ia\textrm{Im}T/M_{pl}}-(1+ax/3)e^{-ax/2}),
\end{eqnarray}
 where we have used $e^{-aT} = e^{-\frac{a}{2}(X+\Phi/3M_{pl}^2)}e^{-a i\textrm{Im}T/M_{pl}}$.

Then, using
\begin{eqnarray}
W^h_T = -\frac{1}{M_{pl}}aAe^{-a(X+\Phi/3M_{pl}^2)/2}e^{-ia\textrm{Im}T/M_{pl}},\quad |W^h_T|^2 = \frac{1}{M_{pl}^2}a^2A^2e^{-a(X+\Phi/3M_{pl}^2)},
\end{eqnarray}
we obtain
\begin{eqnarray}
V_{soft} &\equiv&  
\frac{aAe^{-a(X+\Phi/3M_{pl}^2)/2}}{M_{pl}^2X^{2}}[W^oe^{ia\textrm{Im}T/M_{pl}}+\bar{W}^oe^{-ia\textrm{Im}T/M_{pl}}]
+\frac{1}{9}\frac{a^2A^2e^{-a(X+\Phi/3M_{pl}^2)}}{M_{pl}^4X^{2}}\Phi_{i}\Phi^{i\bar{j}}\Phi_{\bar{j}} \nonumber \\ &&
- \frac{1}{3}\frac{aAe^{-a(X+\Phi/3M_{pl}^2)/2}}{M_{pl}^2X^{2}}
[e^{-ia\textrm{Im}T/M_{pl}}\Phi_{i}\Phi^{i\bar{j}}\bar{W}^o_{\bar{j}}
+e^{ia\textrm{Im}T/M_{pl}}W^o_{i}\Phi^{i\bar{j}}\Phi_{\bar{j}}] 
+\frac{1}{X^{2}}W^o_i\Phi^{i\bar{j}}\bar{W}^o_{\bar{j}}. \nonumber  \\ &&
\end{eqnarray}

At the true vacuum we have $a\textrm{Im}T/M_{pl}=n\pi, ~ X=x,~z^I=0$ where $n$ is an even integer, so the soft terms become
\begin{eqnarray}
V_{soft} &\equiv&  
\frac{aAe^{-ax/2}}{M_{pl}^2x^{2}}[W^o+\bar{W}^o]
+\frac{1}{9}\frac{a^2A^2e^{-ax}}{M_{pl}^4x^{2}}\Phi_{i}\Phi^{i\bar{j}}\Phi_{\bar{j}} \nonumber\\&&
- \frac{1}{3}\frac{aAe^{-ax/2}}{M_{pl}^2x^{2}}
[\Phi_{i}\Phi^{i\bar{j}}\bar{W}^o_{\bar{j}}
+W^o_{i}\Phi^{i\bar{j}}\Phi_{\bar{j}}] 
+\frac{1}{x^{2}}W^o_i\Phi^{i\bar{j}}\bar{W}^o_{\bar{j}}.
\end{eqnarray}
From $A = \sqrt{\frac{3x(M_I^4-\Lambda)e^{ax}}{a^2}}M_{pl} \approx \frac{\sqrt{3}}{a}x^{1/2}e^{ax/2}M_I^2M_{pl}$, we find $aAe^{-ax/2} = \sqrt{3}x^{1/2}M_I^2M_{pl}$. Inserting this expression into the soft-terms potential, we get
\begin{eqnarray}
V_{soft} &\equiv&  
\frac{\sqrt{3}x^{1/2}M_I^2M_{pl}}{M_{pl}^2x^{2}}[W^o+\bar{W}^o]
+\frac{1}{9}\frac{(\sqrt{3}x^{1/2}M_I^2M_{pl})^2}{M_{pl}^4x^{2}}\Phi_{i}\Phi^{i\bar{j}}\Phi_{\bar{j}} \nonumber\\&&
- \frac{1}{3}\frac{\sqrt{3}x^{1/2}M_I^2M_{pl}}{M_{pl}^2x^{2}}
[\Phi_{i}\Phi^{i\bar{j}}\bar{W}^o_{\bar{j}}
+W^o_{i}\Phi^{i\bar{j}}\Phi_{\bar{j}}] 
+\frac{1}{x^{2}}W^o_i\Phi^{i\bar{j}}\bar{W}^o_{\bar{j}}.
\end{eqnarray}
The soft-terms potential thus reduces to
\begin{eqnarray}
V_{soft} &\equiv&  
\frac{\sqrt{3}x^{-3/2}M_I^2}{M_{pl}}[W^o+\bar{W}^o]
+\frac{1}{3}\frac{M_I^4}{M_{pl}^2x}\Phi_{i}\Phi^{i\bar{j}}\Phi_{\bar{j}} \nonumber\\&& 
- \frac{1}{\sqrt{3}}\frac{x^{-3/2}M_I^2}{M_{pl}}  
[\Phi_{i}\Phi^{i\bar{j}}\bar{W}^o_{\bar{j}}  
+W^o_{i}\Phi^{i\bar{j}}\Phi_{\bar{j}}] 
+\frac{1}{x^{2}}W^o_i\Phi^{i\bar{j}}\bar{W}^o_{\bar{j}}. 
\end{eqnarray}

Next, let us consider a general expansion of the observable-sector superpotential $W^o$
\begin{eqnarray}
W^o(z^i) =\sum_{n=0} \frac{W^o_{i\cdots k}}{n!}  z^i\cdots z^k  = B_0 + S_iz^i + M_{ij}z^iz^j + Y_{ijk}z^iz^jz^k + \cdots, \label{ossp}
\end{eqnarray}
where $W^o_{i\cdots k} \equiv \partial^n W^o(z^i)/\partial z^i \cdots \partial z^k$ and $B_0,S_i,M_{ij},Y_{ijk}$ are constant parameters
determining masses and interactions.

We won't perform a full analysis of all possible ranges of values
for $B_0$, $S_i$, $M_{ij}$ and $Y_{ijk}$; instead, we will simplify out analysis by setting $S_i=0$, so that 
the vacuum of the observable sector is at $z^i=0$, assume that for all $i,j,k$ all $M_{ij}$ and 
$Y_{ijk}$ are of the same order, and set $B_0=0$.
The soft terms in the scalar potential are then generated only by following terms in the expansion of $W^{o}$~\cite{softSUSYbreaking}.
\begin{eqnarray}
W^o(z^i) = M_{ij}z^iz^j + Y_{ijk}z^iz^jz^k,
\end{eqnarray}
where the $M_{ij}$ have mass dimension one and the $Y_{ijk}$ are dimensionless.

This choice also implies that such superpotential does not significantly change the cosmological constant because all 
the minima of the $z^i$ are located at zero. We will choose  the U(1) gauge-invariant K\"ahler function of matter fields as follows
\begin{eqnarray}
\Phi =  \delta_{I\bar{J}}z^{I}\bar{z}^{\bar{J}}+ \delta_{i\bar{j}}z^{i}\bar{z}^{\bar{j}},
\end{eqnarray}
where the first (second) term corresponds to hidden (observable) sector. 

With our simplifying assumptions we obtain
\begin{eqnarray}
V_{soft} &=& \frac{2\sqrt{3}}{3} \frac{x^{-3/2}M_{I}^2}{M_{pl}}\Big[ (M_{ij}z^iz^j+Y_{ijk}z^iz^jz^k) +c.c. \Big] + \frac{M_I^4x^{-1}}{3M_{pl}^2}\delta_{i\bar{j}}z^i\bar{z}^{\bar{j}} \nonumber\\
&&+x^{-2}[ M_{ij}z^j+Y_{ijk}z^jz^k]\delta^{i\bar{j}}[\bar{M}_{\bar{i}\bar{j}}\bar{z}^{\bar{i}}+\bar{Y}_{\bar{i}\bar{j}\bar{k}}\bar{z}^{\bar{i}}\bar{z}^{\bar{k}}]. \label{V_soft}
\end{eqnarray}

We also find the magnitude of the corresponding soft parameters from Eq.~\eqref{V_soft} as 
\begin{eqnarray}
&& \frac{2\sqrt{3}}{3}\frac{M_I^2}{M_{pl}}x^{-3/2}|M_{ij} | \equiv m_{s1}^2, \quad \frac{2\sqrt{3}}{3}\frac{M_I^2}{M_{pl}}x^{-3/2}|Y_{ijk}| \equiv m_{s2}, \quad \frac{1}{3}\frac{M_I^4}{M_{pl}^2}x^{-1} \equiv m_{s3}^2, \nonumber\\
&& |M_{ij}|^2x^{-2} \equiv m_{s4}^2, \quad |Y_{ijk}|^2x^{-2} \equiv \frac{m_{s5}}{M_{pl}}, \quad |M_{ij}||Y_{ijk}|x^{-2} \equiv m_{s6}.
\end{eqnarray}
We observe that during inflation (for large $X$ or $\phi$) all the soft mass parameters are very small. Also, the above result give us the following relations
\begin{eqnarray}
&& x = \frac{M_I^4}{3m_{s3}^2M_{pl}^2} = \frac{H^2}{3m_{s3}^2},\quad| M_{ij}| =\frac{1}{6} \frac{M_I^4}{M_{pl}^2}\frac{m_{s1}^2}{m_{s3}^3}= \frac{1}{6}H^2 \frac{m_{s1}^2}{m_{s3}^3} = \frac{\sqrt{3}}{2}\frac{m_{s1}^2}{H}x^{3/2}, \quad |Y_{ijk}| =\frac{1}{6} \frac{M_I^4}{M_{pl}^2}\frac{m_{s2}}{m_{s3}^3}, \nonumber\\
&& m_{s4} = \frac{1}{4} \frac{m_{s1}^2}{m_{s3}},\quad m_{s5} =\frac{1}{4}  \left(\frac{m_{s2}}{m_{s3}}\right)^2M_{pl}, \quad m_{s6} = \frac{1}{4} \left(\frac{m_{s1}}{m_{s3}}\right)^2m_{s2}. \label{softmasses}
\end{eqnarray}
We note that only $m_{s1},m_{s2},m_{s3}$ are free parameters. However, when we examine the kinetic term in Eq.~\eqref{kinetic_term}
 we observe that at $x=1$ the kinetic terms of the matter multiplets are canonically normalized. The condition $x=1$ then gives 
 $m_{s3} = \frac{H}{\sqrt{3}} = 10^{-6}M_{pl} \sim m_{3/2}$. So in this case, the free parameters reduce to $m_1$ and $m_2$ only. 
Notice that in the regime $m_{s3} \sim m_{3/2}$, the parameter $m_{s1}$ determines the magnitude of $|M_{ij}|$ and $m_{s4}$, 
while $m_{s2}$ determines that of $|Y_{ijk}|$, $m_{s5}$, and $m_{s6}$.

%Notice that one soft mass is of the same order as the super-EeV-scale gravitino mass, {\it i.e.} $m_{3/2}\sim O(10^3)~\textrm{EeV}$. The other soft masses are then identified as follows:
%\begin{eqnarray}
%    && m_{s1} \leq 10^{-15/2}\sqrt{HM_{pl}} = 10^{-10}M_{pl},
%    \quad m_{s2} \leq 10^{-15/2} H = 10^{-25/2}M_{pl}, \nonumber\\
%    && m_{s4},m_{s5}, \leq 10^{-15}M_{pl}, \quad m_{s6} \leq 10^{-45/2} M_{pl}. \label{Small_soft_mass_parameters}
%\end{eqnarray} 
%So we can obtain a weak hierarchy of mass scales in the observable sector, which is given by
%\begin{eqnarray}
%    m_{s6} \lesssim m_{s4},m_{s5} \lesssim m_{s2} < m_{s1} \lesssim m_{s3} \sim m_{3/2}. 
%\end{eqnarray}
%Notice that except for $m_{s3} \sim m_{3/2}$, the other mass scales can be arbitrarily smaller than $m_{s3}$. The precise hierarchy of these mass scales will be determined after specifying each of the soft mass parameters. The key fact here is that there exist very small mass parameters, which enable matter scalars to be lighter than the gravitino.

Finally, let us investigate further 
the physical masses of matter scalars in the observable sector. Here, we are going to look only at the matter scalar masses and leave a detailed study of fermion masses and interactions to a future work,
since the purpose of this section is to demonstrate the existence of light scalars in the observable sector, 
whose masses can be smaller than that of the gravitino. Because of the soft mass parameters we found, we expect that some scalars will be as heavy as the gravitino, while other scalars could be much  lighter.

To compute the scalar masses we must remember to include contributions coming from the expansion of the hidden-sector 
potential to second order in the observable-sector scalars $z^i$: $V_h(z^I,z^i)=V_h(0.0) + V_{h\; i{}\bar{j}} z^i {\bar{z}}^{{}\bar{j}}$. We thus consider the general expression for the total scalar potential, which is written with the canonical mass dimensions by 
\begin{eqnarray}
V &=& V_D+V_F^h + V_{soft} 
\nonumber\\
&=& \frac{1}{2}g^2  M_{pl}^4 \Big(\xi+\frac{q_Iz^I\Phi_I+q_I\bar{z}^{\bar{I}}\Phi_{\bar{I}}}{XM_{pl}^2 }\Big)^2 \nonumber\\
&&- \frac{1}{X^2M_{pl}^2}\bigg(-2aA^2e^{-a(X+\Phi/3M_{pl}^2)}+2acA^2 e^{-a(X+\Phi/3M_{pl}^2)/2} \cos(a\textrm{Im}T/M_{pl})\bigg)
\nonumber\\
&&+ \frac{1}{3X^{2}M_{pl}^2}\Big(X+\frac{1}{3M_{pl}^2}\Phi_I\Phi^{I\bar{J}}\Phi_{\bar{J}}\Big) a^2A^2 e^{-a(X+\Phi/3M_{pl}^2)}
\nonumber\\
&&+\frac{aAe^{-a(X+\Phi/3M_{pl}^2)/2}}{M_{pl}^2X^{2}}[W^oe^{ia\textrm{Im}T/M_{pl}}+\bar{W}^oe^{-ia\textrm{Im}T/M_{pl}}]  +\frac{1}{9}\frac{a^2A^2e^{-a(X+\Phi/3M_{pl}^2)}}{M_{pl}^4X^{2}}\Phi_{i}\Phi^{i\bar{j}}\Phi_{\bar{j}} \nonumber \\ &&
- \frac{1}{3}\frac{aAe^{-a(X+\Phi/3M_{pl}^2)/2}}{M_{pl}^2X^{2}}
[e^{-ia\textrm{Im}T/M_{pl}}\Phi_{i}\Phi^{i\bar{j}}\bar{W}^o_{\bar{j}}
+e^{ia\textrm{Im}T/M_{pl}}W^o_{i}\Phi^{i\bar{j}}\Phi_{\bar{j}}] 
+\frac{1}{X^{2}}W^o_i\Phi^{i\bar{j}}\bar{W}^o_{\bar{j}}.
\end{eqnarray}

First, we find that masses of the hidden-sector matter scalars $z^I$ and $\textrm{Im}T$ can be independently defined by tuning the magnitude of the U(1) gauge charge $\forall I:q_I \equiv q$ and the parameter $a$ respectively such that they are positive definite. This implies that the hidden-sector fields can be heavy as much as we wish. Thus, to get an effective single-field slow-roll inflation we should make the hidden-sector matter scalars much heavier than the Hubble scale during slow roll. Their masses can be lighter than the Hubble scale before the onset of the slow-roll period, that is for
very large values of $X$. Second, it is obvious that the inflaton mass is of the same order as the Hubble scale, i.e. $m_{\phi} \sim H$, since the scalar potential is of ``HI'' or ``Starobinsky''form and has a de Sitter vacuum, as we have
seen in the previous sections.

Next, we investigate masses of the observable-sector fields. We can simplify further our analysis to make our point 
clearer by assuming that the quadratic term in the superpotential is diagonal $M_{ij}=\delta_{ij}M$. From the total scalar potential, we find the observable-sector squared mass matrix $M_{obs}^2$ at the vacuum specified by the conditions that $a\textrm{Im}T=n\pi$, $z^I=0$, and $z^i=0$
\begin{eqnarray}
M^2_{obs} \equiv \begin{pmatrix}
V_{i\bar{j}} & V_{ij} \\ 
V_{\bar{i}\bar{j}} & V_{\bar{i}j} 
\end{pmatrix} 
\end{eqnarray}
where
\begin{eqnarray}
V_{i\bar{j}} &=& -\frac{2a^2A^2}{3X^2}\delta_{i\bar{j}}e^{-aX} 
+ \frac{ca^2A^2}{3X^2}\delta_{i\bar{j}}e^{-aX/2}  - \frac{a^3A^2}{9X}e^{-aX} \delta_{i\bar{j}} + \frac{a^2A^2}{9X^2}e^{-aX}\delta_{i\bar{j}} 
+\frac{1}{X^2} W_{il}^o\Phi^{l\bar{n}}\bar{W}_{\bar{n}\bar{j}}^o \nonumber\\
&=& \frac{1}{X^2} W_{il}^o\Phi^{l\bar{n}}\bar{W}_{\bar{n}\bar{j}}^o -\frac{2a^2A^2}{9X^2}e^{-aX}\delta_{i\bar{j}} = \Big(\frac{M^2}{X^2}-\frac{2a^2A^2}{9X^2}e^{-aX}\Big)\delta_{i\bar{j}},
\nonumber\\
V_{ij} &=& \frac{aA}{3X^2}e^{-aX/2}W_{ij}^o  = \frac{aA}{3X^2}e^{-aX/2} M \delta_{ij}. 
\end{eqnarray}

Restoring the mass dimension, the mass eigenvalues are 
\begin{eqnarray}
    m_{\pm}^2 &\equiv& \Big(\frac{M^2}{X^2}-\frac{2a^2A^2}{9X^2M_{pl}^4}e^{-aX}\Big)  \pm \frac{aA}{3X^2M_{pl}^2}e^{-aX/2} M  \nonumber\\
    &=& \frac{1}{X^2}\left(M\pm \frac{aA}{6M_{pl}^2}e^{-aX/2}\right)^2 -\frac{a^2A^2}{4X^2M_{pl}^4}e^{-aX}. \label{GeneralScalarMass}
\end{eqnarray}
We observe that if $M \sim aA/M_{pl}^2$ (which is equivalent to the condition that $m_{s1} \sim H$), then both masses $m_{\pm}$ are positive definite for all $X=e^{\sqrt{2/3}\phi}>0$ (or all $\phi$), which means that during inflation the matter scalar masses are well defined (and become very light for large values of $X$ or $\phi$). This will be confirmed in the following.

Let us check the values of the scalar masses on the post-inflationary vacuum. Using the relations $A=M_{pl}\sqrt{\frac{3x(M_I^4-\Lambda)e^{ax}}{a^2}}$, $c=(1+\frac{ax}{3})e^{-ax/2}$, and $M = \frac{\sqrt{3}}{2}\frac{m_{s1}^2}{H}x^{3/2}$ in Eq. \eqref{softmasses} and setting $\Lambda \approx 0$ at $X=x=1$, where the kinetic terms of the matter scalars are canonically normalized, we obtain 
\begin{eqnarray}
      m_{\pm}^2 = \frac{3}{4}\frac{M_{pl}^2}{M_{I}^4}m^4_{s1} -\frac{2}{3}\frac{M_I^4}{M_{pl}^2} \pm \frac{1}{2}m_{s1}^2 = \Big(\frac{3}{4}k^2 \pm k - \frac{2}{3}\Big)H^2,\label{MassEigenvalues}
\end{eqnarray}
in which we define $m_{s1}^2 \equiv k H^2$ (where $k>0$ and $M_I^2/M_{pl} = H$). We notice that physical masses of scalars are determined only by the ``free'' parameter $m_{s1}$ (or $k$) and the Hubble mass $H$. Positivity of the physical masses ``$m_\pm$'' imposes the inequality
\begin{eqnarray}
m_+:~-\frac{2}{3}+\frac{2}{3}\sqrt{3} <  k, \quad     m_-:~\frac{2}{3}+\frac{2}{3}\sqrt{3} <  k \implies \frac{2}{3}+\frac{2}{3}\sqrt{3} <  k \label{LightScalarCondition}
\end{eqnarray}
Then, with this inequality, we can choose an arbitrary value of $k$ such that
\begin{eqnarray}
 \frac{2}{3}+\frac{2}{3}\sqrt{3} <    k = \frac{2}{3} + \frac{2}{3}\sqrt{3(1+m_-^2/H^2)} = -\frac{2}{3} + \frac{2}{3}\sqrt{3(1+m_+^2/H^2)}
\end{eqnarray}
allowing one physical mass $m_-$ to be parametrically lighter than the other physical mass $m_+$ as
\begin{eqnarray}
m_+^2 = \frac{4H^2\left(1+\sqrt{3(1+m_-^2/H^2)}\right)+3m_-^2}{3}.
\end{eqnarray}
We note that $m_{s1} = \sqrt{k}H \sim H$ when $m_- \ll H$, implying that $m_{\pm}^2 >0$ is indeed satisfied. In this limit, we find that one physical mass $m_-$ can be much smaller than the Hubble scale, while the other physical mass are of the order of the Hubble scale: 
\begin{eqnarray}
  m_- \ll H, \quad m_+ \gtrsim  H.
\end{eqnarray}

From this we note that in the observable sector after inflation (that is at $x=1$) one physical mass $m_-$ can be lighter than that of the gravitino, while the other physical mass $m_+$ becomes of the same order of the gravitino mass. 
We also note that the matter scalar with 
masses of order of the super-EeV gravitino mass ($\sim 10^{-6}M_{pl}$) may be a candidate for heavy dark matter candidate, because
it is in the mass range $10^{-8}M_{pl} \leq m_{\chi} \leq M_{pl}$, which is outside the excluded region shown in Figs. 2, 3, and 4 of 
ref.~\cite{HDM}.  
To summarize, we found the following constraints on soft masses. First, $m_{s1}$ must satisfy Eq.~\eqref{LightScalarCondition} to 
allow for some light scalars while $m_{s3}$ is of the same order as the gravitino mass $m_{3/2}$ and $m_{s4}$ is determined by the 
chosen value of $m_{s1}$. Notice that all these mass parameters are subject to strict constraints such as Eq.~\eqref{LightScalarCondition}. Furthermore, $m_{s2}$, $m_{s5}$, and $m_{s6}$ can be arbitrarily small, $m_{s5}$ and $m_{s6}$ are proportional to $m_{s2}^2$ and $m_{s2}$ respectively, and $m_{s2}$ is a free parameter.

It is worth noticing that the observable sector masses $m_-$ are compatible with the ``Case 1'' reheating-scenario condition of ref.~\cite{LightScalar}, for which single-field plateau-potential inflation is robust under the introduction of light scalars. 
The parameters characterizing the reheating scenario are
\begin{eqnarray}
    \Gamma_{\phi} < \Gamma_{z^i} < m_{z^i} \sim m_{-} < H , \quad \frac{\left<z^i\right>}{M_{pl}}  \ll 1,
\end{eqnarray}
where $\Gamma_{\phi},\Gamma_{z^i}$ are the decay rates of $\phi$ and $z^i$ during the reheating phase and $\left<z^i\right>$ are the expectation values of matter scalars $z^i$ after inflation. We note that $\frac{\left<z^i\right>}{M_{pl}}  \ll 1$ implies that the slow-roll inflation should begin around the minima of matter scalars, so that at the end of inflation the corresponding vacuum expectation values will be much smaller than the Planck scale $M_{pl}$. Hence, as long as the above ``Case 1'' reheating-scenario condition is satisfied, the slow-roll inflation in our model will effectively be driven by a single inflaton field $\phi$ along the minima of the matter scalars.

\section{Conclusions and outlook}
To summarize our findings, we have seen that our model can naturally produce plateau-potential inflation at the Hubble scale with a high scale spontaneously supersymmetry breaking in the hidden sector and low scale soft supersymmetry breaking interactions with various soft 
masses in the observable sector. We also obtain naturally a super-EeV gravitino, which is compatible with  
constraint for heavy gravitino cold dark matter ({\it i.e.} $0.1 ~\textrm{EeV} \lesssim m_{3/2} \lesssim 1000~\textrm{EeV}$) 
\cite{Inf_HighSUSY_EeVGrav}. In this work, we have not investigated the specific structure and dynamics of observable-sector 
interactions or the detailed construction of a realistic low energy effective theory of the observable sector. It would be of obvious interest
to see how far this scenario could be pursued and how to incorporate in it a supersymmetric extension of the Standard Model or a
Grand Unified Theory. Models with a dynamically generated FI term, realistic observable sector, 
D-term inflation, and high-scale supersymmetry breaking have been studied in~\cite{DS17}; other uses of FI terms for inflation were 
presented in~\cite{li14}. It would be interesting to reproduce the phenomenologically desirable features of~\cite{DS17,li14}  and other 
models proposed in the literature in our scenario.
On a different note, it would be extremely interesting to see if the new FI term in general and in our KKLT-type
scenario in particular 
can be obtained in string theory. In other words, it would be interesting to see if our model belongs to the string landscape or
the string swampland \cite{Swampland} (See refs. \cite{Recent_Review_Swampland1,Recent_Review_Swampland2} for recent reviews of  swampland conjectures). Here we shall just mention that the magnetic Weak Gravity Conjecture given e.g. in eq.~(3.7) of 
ref.~\cite{Recent_Review_Swampland2} can be easily satisfied in our model.

On a more concrete note, a detailed analysis and derivation of the bounds on the new FI terms that were crucial in this paper 
is highly desirable. This will be the main focus of our forthcoming paper~\cite{jp20}.

\subsection*{Acknowledgments} 
  M.P.\ is supported in part by NSF grant PHY-1915219.

\appendix

\end{document}